\documentclass[sigconf]{acmart}

\graphicspath{{figure/}{figures/}}

\newcommand{\mf}{\mathbf}
\newcommand{\mc}{\mathcal}
\newcommand{\mb}{\mathbb}

\newcommand{\bs}{\boldsymbol}

\DeclareMathOperator*{\argmax}{argmax}
\setcopyright{acmcopyright}

\usepackage[utf8]{inputenc}
\usepackage{pgfplots}
\usepackage{amsmath}
\usetikzlibrary{patterns}
\acmDOI{10.475/123_4}

\acmISBN{123-4567-24-567/08/06}

\acmConference[MobiHoc'19]{ACM International Symposium on Mobile Ad Hoc Networking and Computing}{July 2019}{Catania, Italy} 
\acmYear{2019}
\copyrightyear{2019}

\acmPrice{15.00}

\begin{document}
\title[DeepRadioID: Real-Time Optimization of Deep Learning-based Radio Fingerprinting Algorithms]{\textit{DeepRadioID}: Real-Time Channel-Resilient Optimization of Deep Learning-based Radio Fingerprinting Algorithms}

\author{Francesco Restuccia}
\affiliation{%
  \institution{Northeastern University}
  \streetaddress{360 Huntington Ave}
  \city{Boston} 
  \state{MA} 
    \country{USA}
  \postcode{02115}
}
\email{frestuc@northeastern.edu}

\author{Salvatore D'Oro}
\affiliation{%
  \institution{Northeastern University}
  \streetaddress{360 Huntington Ave}
  \city{Boston} 
  \state{MA} 
    \country{USA}
  \postcode{02115}
}
\email{s.doro@northeastern.edu}

\author{Amani Al-Shawabka}
\affiliation{%
  \institution{Northeastern University}
  \streetaddress{360 Huntington Ave}
  \city{Boston} 
  \state{MA} 
    \country{USA}
  \postcode{02115}
}
\email{amani@northeastern.edu}

\author{Mauro Belgiovine}
\affiliation{%
  \institution{Northeastern University}
  \streetaddress{360 Huntington Ave}
  \city{Boston} 
  \state{MA} 
    \country{USA}
  \postcode{02115}
}
 \email{belgiovine@northeastern.edu}

\author{Luca Angioloni}
\affiliation{%
  \institution{Northeastern University}
  \streetaddress{360 Huntington Ave}
  \city{Boston} 
  \state{MA} 
    \country{USA}
  \postcode{02115}
}
\email{angioloni@northeastern.edu}

\author{Stratis Ioannidis}
\affiliation{%
  \institution{Northeastern University}
  \streetaddress{360 Huntington Ave}
  \city{Boston} 
  \state{MA} 
    \country{USA}
  \postcode{02115}
}
\email{ioannidis@northeastern.edu}

\author{Kaushik Chowdhury}
\affiliation{%
  \institution{Northeastern University}
  \streetaddress{360 Huntington Ave}
  \city{Boston} 
  \state{MA} 
    \country{USA}
  \postcode{02115}
}
\email{krc@northeastern.edu}

\author{Tommaso Melodia}
\affiliation{%
  \institution{Northeastern University}
  \streetaddress{360 Huntington Ave}
  \city{Boston} 
  \state{MA} 
  \country{USA}
  \postcode{02115}
}
\email{melodia@northeastern.edu}

\renewcommand{\shortauthors}{F.~Restuccia et al.}


\begin{abstract}
Radio fingerprinting provides a reliable and energy-efficient IoT authentication strategy by leveraging the unique hardware-level imperfections imposed on the received wireless signal by the transmitter's radio circuitry. Most of existing approaches utilize hand-tailored protocol-specific feature extraction techniques, which can identify devices operating under a pre-defined wireless protocol only. Conversely, by mapping inputs onto a very large feature space, deep learning algorithms can be trained to fingerprint large populations of devices operating under any wireless standard.

One of the most crucial challenges in radio fingerprinting is to counteract the action of the wireless channel, which decreases fingerprinting accuracy significantly by disrupting hardware impairments. On the other hand, due to their sheer size, deep learning algorithms are hardly re-trainable in real-time. Another aspect that is yet to be investigated is whether an adversary can successfully impersonate another device's fingerprint. To address these key issues, this paper proposes \emph{DeepRadioID}, a system to optimize the accuracy of deep-learning-based radio fingerprinting algorithms \textit{without retraining the underlying deep learning model}. The key intuition is that through the application of a carefully-optimized digital finite input response filter (FIR) at the transmitter's side, we can apply tiny modifications to the waveform to strengthen its fingerprint according to the current channel conditions. We mathematically formulate the \textit{Waveform Optimization Problem} (WOP) as the problem of finding, for a given trained neural network, the optimum FIR to be used by the transmitter to improve its fingerprinting accuracy. 

We extensively evaluate \emph{DeepRadioID} on a experimental testbed of 20 nominally-identical software-defined radios, as well as on two datasets made up by 500 ADS-B devices and by 500 WiFi devices provided by the DARPA RFMLS program. Experimental results show that \emph{DeepRadioID} (i) increases fingerprinting accuracy by about 35\%, 50\% and 58\% on the three scenarios considered; (ii) decreases an adversary's accuracy by about 54\% when trying to imitate other device's fingerprints by using their filters; (iii) achieves 27\% improvement over the state of the art on a 100-device dataset.
\end{abstract}



\maketitle



%




\section{Introduction}


Thanks to its unprecedented pervasiveness, one the most crucial issues in the Internet of Things (IoT) is designing \emph{scalable, reliable and energy-efficient authentication mechanisms} \cite{zhao2013survey,Restuccia-IoT2018}. However, most of the existing authentication mechanisms are not well-suited to the IoT since they are heavily based on cryptography-based algorithms and protocols, which are often too computational expensive to be run on tiny, energy-constrained IoT devices \cite{sicari-2015security}. 

To address this key issue, a number of techniques based on radio fingerprinting have been proposed over the last few years \cite{10_brik2008wireless,8_nguyen2011device,12_vo2016fingerprinting,16-Peng-ieeeiotj2018,17-Xie-ieeeiotj2018,18-Xing-ieeecomlet2018}. The core intuition behind radio fingerprinting is that wireless devices usually suffer from small-scale hardware-level imperfections typically found in off-the-shelf RF circuitry, such as phase noise, I/Q imbalance, frequency and sampling offset, and harmonic distortions \cite{johnson1966physical}. We can thus obtain a ``fingerprint'' of a wireless device by estimating the RF impairments on the received waveform and associating them to a given device  \cite{1_xu2016device}. 

Traditional techniques for radio fingerprinting (which are discussed in details in Section \ref{sec:rw}) rely on complex feature-extraction techniques that leverage protocol-specific characteristics (such as WiFi pilots/training symbols \cite{10_brik2008wireless,12_vo2016fingerprinting} or ZigBee O-QPSK modulation \cite{16-Peng-ieeeiotj2018}) to extract hardware impairments. Therefore, they are not general-purpose in nature and are hardly applicable to the IoT, where a plethora of different wireless protocols are used \cite{zorzi2010today}. To overcome this limitation, in this paper we use techniques based on \textit{deep learning} \cite{jagannath2019machine} to design \textit{general-purpose, high-performance, and optimizable} radio fingerprinting algorithms. Thanks to the very large number of parameters (\textit{i.e.}, in the order of $10^6$ or more), deep neural networks can analyze unprocessed I/Q samples without the need of application-specific and computational-expensive feature extraction and selection algorithms \cite{Restuccia-infocom2019}. \vspace{0.1cm} 

\textbf{Challenges.}~There are a number of critical issues in applying deep learning techniques to RF fingerprinting. First, deep learning models usually require a significant time to be re-trained, even with modern GPUs \cite{chetlur2014cudnn}. Therefore, we cannot assume that \textit{the underlying deep learning model can be retrained in real time}. Second, a fingerprinting system must evaluate \textit{the impact of adversarial actions}. Specifically, to the best of our knowledge, existing work has not yet evaluated if and when an adversary can imitate a legitimate device's fingerprint. Last, but not least, we need to \textit{address the (potentially disruptive) action of the wireless channel on the system's fingerprinting accuracy}. This is because, due to channel action, two identical waveforms transmitted by the same RF interface at two different moments in time are usually different from each other. This implies that the models will operate on \textit{non-stationary} input data \cite{goodfellow2016deep}, which significantly decreases the model's fingerprinting accuracy when the classifier is used on data collected with a wireless channel that is significantly different from the one used to train it.

\begin{figure}[!h]
    \centering
    \includegraphics[width=\columnwidth]{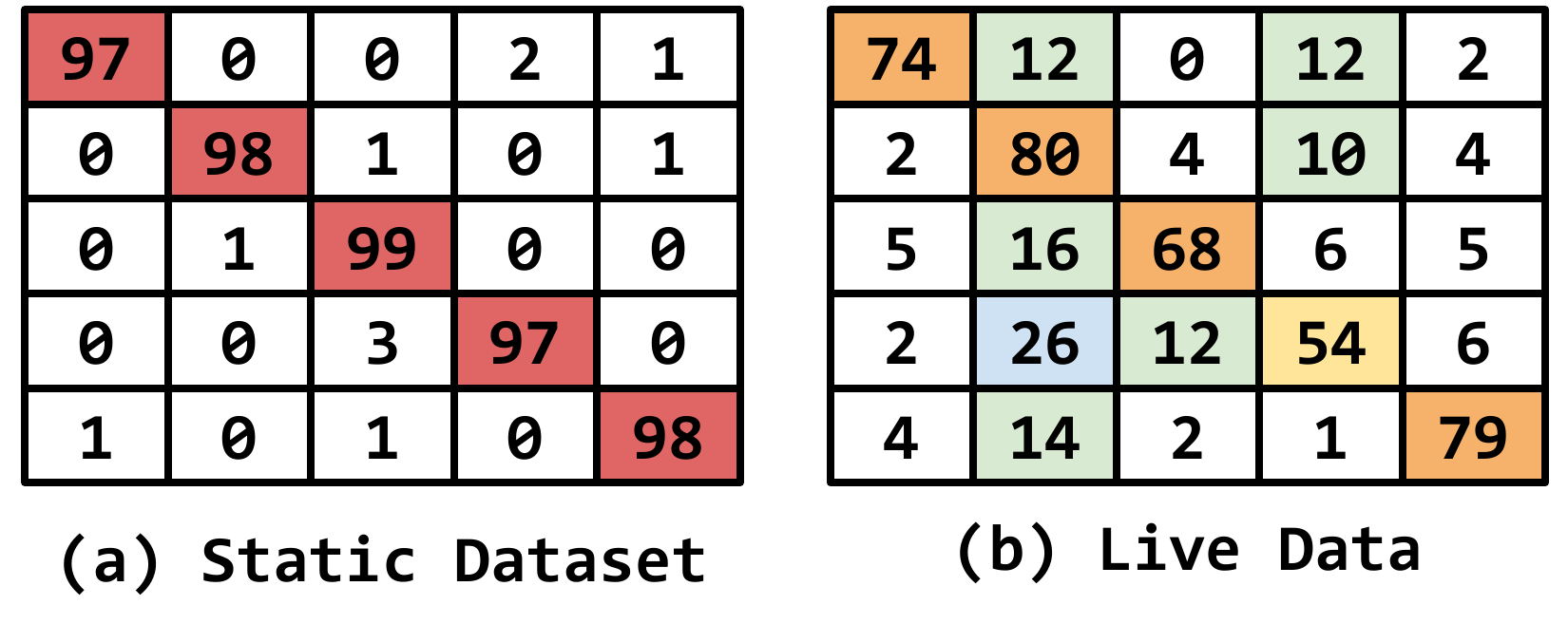}
    \caption{Confusion matrices of 5-device bit-similar model with (a) original dataset; (b) live-collected data. The figure highlights that different wireless channel conditions imply a loss in fingerprinting accuracy.\vspace{-0.5cm}}
    \label{fig:conf_mat_intro}
\end{figure}

To illustrate this crucial point, Figure \ref{fig:conf_mat_intro} shows the confusion matrices of a deep learning model trained to fingerprint 5 devices through the experimental testbed that will be presented in Section \ref{sec:exp_res}. The confusion matrix (a) was computed on data collected approximately 5 minutes after the training data was collected, while Figure \ref{fig:conf_mat_intro}(b) was obtained by testing the model on completely new data collected 7 days after the model was trained. Figure \ref{fig:conf_mat_intro} remarks that the fingerprinting accuracy decreases significantly when data collected under completely different channel conditions is fed to the model, demonstrating that the channel's action must indeed be addressed through real-time optimization.\vspace{0.1cm}

\textbf{Novel Contributions.}~This paper addresses the above challenges by making the following novel contributions:

$\bullet$ We propose \emph{DeepRadioID}, a system for real-time channel- and adversary-resilient optimization of deep-learning-based radio fingerprinting algorithms. The key innovation behind \emph{DeepRadioID} is to leverage a carefully-optimized digital finite input response filter (FIR) at the transmitter's side, which slightly modifies its baseband signal to compensate for current channel conditions. The optimal FIR is computed by the receiver and sent back as feedback to the transmitter. We postulate the \textit{Waveform Optimization Problem} (WOP) to find the optimal FIR, and derive a novel algorithm based on the \textit{Nonlinear Conjugate Gradient} (NCG) method  to efficiently solve it. We show in Section \ref{sec:ber} that the FIR's action can be effectively compensated at the receiver's side through the discrete Fourier transform (DFT) of the received signal, thus causing a negligible throughput decrease (\textit{i.e.}, less than 0.2\% in our experiments).

$\bullet$ We extensively evaluate the performance of \emph{DeepRadioID} on an experimental testbed made up of 20 bit-similar devices (\textit{i.e}., transmitting the same baseband signal through nominally-identical RF interfaces and antennas). To evaluate the scalability of \emph{DeepRadioID} and to experiment with different wireless technologies and deeper learning models, we also leverage two datasets of IEEE 802.11a/g (WiFi) and Automatic Dependent Surveillance -- Broadcast (ADS-B) transmissions, each containing 500 devices. These transmissions were collected ``in the wild'' by DARPA for the RFMLS program. To the best of our knowledge, we are the first ever to evaluate radio fingerprinting algorithms on datasets of such dimension. Experimental results indicate that (i) an adversary trying to imitate a legitimate device's fingerprint by using the same FIR filter \textit{decreases} its fingerprinting accuracy by about 54\% on the average; (ii) DeepRadioID \textit{increases} the fingerprinting accuracy by (a) 35\% on our bit-similar experimental testbed, and (b) by 50\% and by 58\% on the 500-device ADS-B and WiFi datasets, respectively; (iii) by comparing with the state of the art \cite{12_vo2016fingerprinting}, \emph{DeepRadioID} improves the accuracy by about 27\% on a reduced dataset of 100 WiFi devices.\vspace{-0.2cm} 

\section{Related Work}\label{sec:rw}

Radio fingerprinting has received significant attention over the last few years -- for an excellent survey paper on the topic, the reader may refer to \cite{1_xu2016device}. 

The vast majority of existing work has applied carefully-tailored feature extraction techniques at the physical layer to fingerprint wireless devices
\cite{10_brik2008wireless,8_nguyen2011device,12_vo2016fingerprinting,16-Peng-ieeeiotj2018,17-Xie-ieeeiotj2018,18-Xing-ieeecomlet2018}. Nguyen \emph{et al.}~\cite{8_nguyen2011device} use device-dependent channel-invariant radio-metrics and propose a non-parametric Bayesian method to detect the number of devices. However, the effectiveness of the features is proven with a testbed made up of only four ZigBee transmitters. Brik \emph{et al.}~\cite{10_brik2008wireless} considered a combination of frequency offset, transients, and constellation errors to fingerprint 130 IEEE~802.11b cards with an accuracy of 99\%. However, conversely from ours, the experiments in \cite{10_brik2008wireless} were performed in an RF-insulated environment (\textit{i.e.}, without any channel effect), thus the algorithms' effectiveness in real-world environments has yet to be established. More recently, Vo \emph{et al.} \cite{12_vo2016fingerprinting} proposed a series of algorithms with features based on frequency offsets, transients and the WiFi scrambling seed, and validated them with off-the-shelf WiFi cards in a non-controlled RF environment, achieving accuracy between 44 and 50\% on 93 devices. In Section \ref{sec:dataset_res}, we show that \emph{DeepRadioID} improves the acccuracy of 27\% on a 100-device WiFi testbed. Recently, Peng \textit{et al.}~\cite{16-Peng-ieeeiotj2018} proposed fingerprinting algorithms for ZigBee devices  based on modulation-specific features such as differential constellation trace figure (DCTF), showing that their features achieve almost 95\% accuracy on a 54-radio testbed. 

The key drawback of existing feature-based fingerprinting techniques is that they are inherently tailored for a specific wireless technology (\textit{e.g.}, WiFi or ZigBee), which ultimately limits their applicability to IoT scenarios where devices operate under different standards \cite{zorzi2010today}. Moreover, existing work on feature-based fingerprinting has not considered the problem of optimizing the algorithm's accuracy in real-time. Thus, we consider \emph{deep learning} to design a general-purpose and scalable fingerprinting system. Although watermarking has been proposed to identify devices in the IoT \cite{ferdowsi2018deep}, it requires the insertion of additional information, which \emph{DeepRadioID} avoids.

The closest work to ours is \cite{5_8466371}, where the authors proposed the usage of convolutional neural networks to fingerprint nominally-identical USRP X310 devices. They also show that by using artificially introduced hardware impairments at the transmitter's side, the accuracy can be improved to 99\%. However, \cite{5_8466371} suffers from the following key limitations: (i) the artificial impairments cannot be accurately compensated at the receiver's side; (ii) the relationship between hardware impairment and accuracy is not fully characterized; and (iii) adversarial action is not considered. \emph{DeepRadioID} overcomes the above limitations by proposing a system where we increase in real-time the fingerprinting accuracy through the application of a FIR filter that (a) is obtained through rigorous optimization (Section \ref{sec:gradient}); (b) can be compensated at the receiver's side (Section \ref{sec:ber}); (c) cannot be used by an adversary to impersonate another device (Section \ref{sec:exp_res}).\vspace{-0.2cm}

\begin{figure*}
  \centering
  \includegraphics[width=\linewidth]{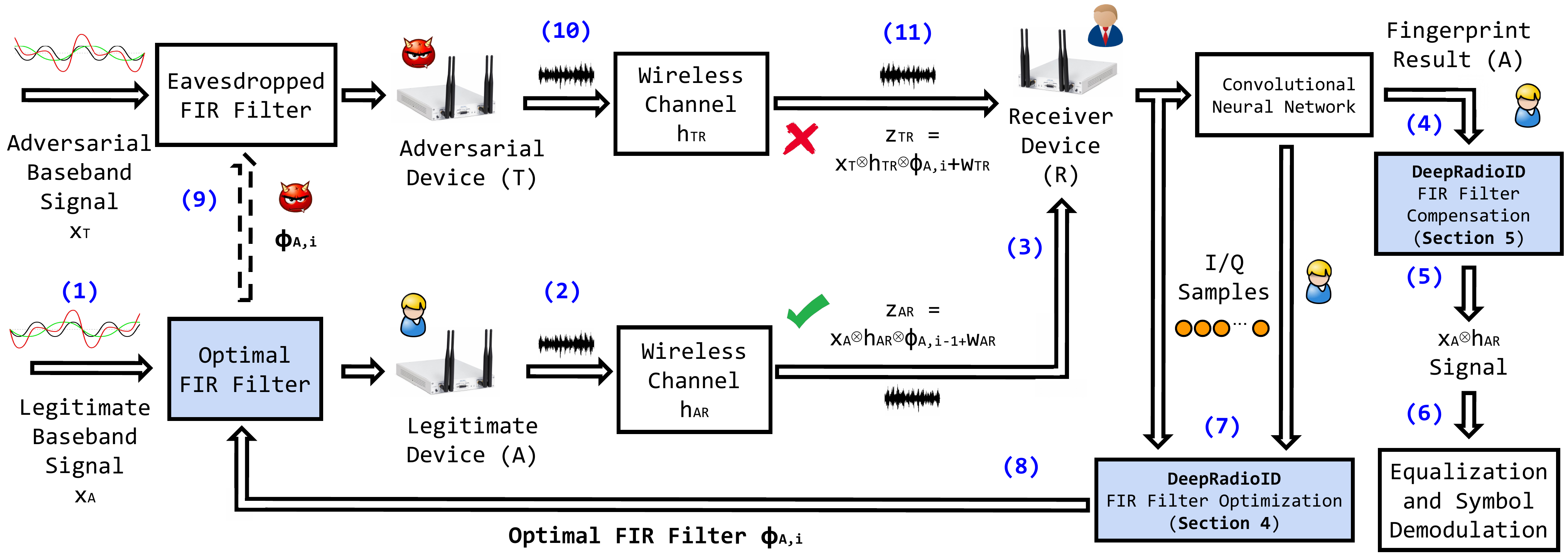}
   \caption{\label{fig:arch}A high-level overview of the \emph{DeepRadioID} system, where we also illustrate an adversary (T) trying to impersonate a legitimate device (A) using an eavesdropped FIR filter. Since A's FIR filter has been tailored to match A's unique channel and impairment conditions, we show in Section \ref{sec:exp_res} that T does not improve its fingerprinting accuracy by using A's filter.}\vspace{-0.4cm}
\end{figure*}

\section{\texorpdfstring{D\MakeLowercase{eep}R\MakeLowercase{adio}ID}{DeepRadioID}: An Overview} \label{sec:system}

We first discuss some key observations and motivations to motivate our design choices in Section \ref{sec:observations}. We then provide an in-depth description and a walk-through of the main steps involved in the fingerprinting process in Section \ref{sec:overview}.

\subsection{DeepRadioID: Key Intuitions}\label{sec:observations}

The need to optimize the accuracy of fingerprinting systems arises from the fact that the wireless channel is dynamic and almost unpredictable in nature. Thus, hardware impairments such as I/Q imbalances, DC offset, phase noise, carrier/sampling offsets, and power amplifier distortions can be disrupted by the channel's action. Moreover, these impairments are also time-varying and dependent on a number of factors, such as local oscillator (LO) frequency \cite{sourour2004frequency} and current temperature of the RF circuitry \cite{johnson1966physical}. These considerations imply that we cannot assume impairments as perfectly stationary -- hence the need for real-time optimization.

To address the non-stationary nature of the problem, our first observation is that convolutional neural networks (CNNs) have shown to be prodigiously suited to recognize complex ``patterns'' in input data \cite{lecun2015deep} -- these patterns are, in our case, the imperfections in the radio hardware. However, a major challenge that still lingers is \textit{how do we optimize the CNN's output for a given device without retraining the CNN itself.} To answer this question, we devise a new approach based on finite impulse response (FIR) filtering of the transmitter's baseband signal to ``restore'' the patterns that are disrupted by the current channel conditions -- thus making the signal ``more recognizable'' to the CNN. We use FIRs because of the following: (i) FIRs are very easy to implement in both hardware and software on almost any wireless device; (ii) the computation complexity of applying a FIR filter of length $m$ to a signal is $\mathcal{O}(m)$ -- thus it is a very efficient algorithm; and most importantly, (iii) its effect on the BER can be almost perfectly compensated at the receiver's side, as shown in Section \ref{sec:ber}.

However, this approach spurs another challenge, which is \textit{how to set the FIR taps in such a way that the fingerprinting accuracy for a given device is maximized}. Our intuition here is to find the FIR that modifies input $x$ so that the resulting $x^*$ signal maximizes the neuron activation correspondent to a given device, as shown in Section \ref{sec:gradient}. We are able to do this efficiently since the layers inside CNNs, although non-linear, are derivable, and thus we can compute the gradient of the output with respect to the FIR taps according to a given input. This way, we can design an optimization strategy that is fundamentally general-purpose in nature. 

\subsection{DeepRadioID: A Walk-Through}\label{sec:overview}

Figure \ref{fig:arch} provides a walk-through of the main building blocks of \textit{DeepRadioID} and the main operations involved in the fingerprinting process. We highlight with a shade of blue the blocks that are added to the normal modulation/demodulation chain as part of \textit{DeepRadioID}. The walk-through also shows how an adversary may try to imitate another device's fingerprint. The detailed explanation of \emph{DeepRadioID}'s main module will be given in Section \ref{sec:filt_design}.

The first step for a legitimate device ``A" that wants to be authenticated by a receiver ``R'' is to filter its baseband signal with FIR $\bs \phi_{A,i-1}$ (step 1), which was obtained at the previous optimization step. FIR $\phi_{A, 0}$ is set to 1 (\textit{i.e.}, no filtering). The filtered signal is then sent to A's RF interface (step 2). By also accounting the effect of the wireless channel, ``R" will receive a baseband signal ${\bf z}_{AR} = {\bf x}_A \circledast \bs \phi_{A,i-1} \circledast \bf{h}_{AR} + {\bf w}_{AR}$, where ${\bf x}_A$ is the transmitted symbol sequence, $\bf{h}_{AR}$ and ${\bf w}_{AR}$ are the fading and noise introduced by the channel, respectively. The I/Q samples of ${\bf z}_{AR}$ are then fed to a CNN to fingerprint the originating device (step 4). The fingerprinting result is then used to compensate the FIR filter $\bs \phi_{A,i-1}$ (step 5, discussed in Section \ref{sec:ber}), so that the resulting signal is then sent to the symbol demodulation logic to recover the application's data (step 6). The I/Q samples and the fingerprinting result are then fed to the \textit{DeepRadioID} FIR Optimization module (step 7, presented in Section \ref{sec:filt_design}). The optimal FIR filter $\bs \phi_{A,i}$ is then sent back to A to improve its fingerprinting accuracy (step 8).

We now examine the case of adversarial action as follows. We assume that an adversarial device ``T" is capable of eavesdropping A's FIR $\bs \phi_{A,i}$. T's target here is to impersonate A by spoofing A's hardware fingerprint (step 9). After T's baseband signal is transmitted and goes through the wireless channel (step 10), the baseband signal received by R will be $\bf{z}_{TR} = \bf{x}_R \circledast \bs \phi_{A,i-1} \circledast \bf{h}_{TR} + \bf{w}_{TR}$, which is then fed to the CNN as input. However, we show in Section \ref{sec:exp_res} that, since A's FIR filter has been optimized for A's unique hardware impairments and A's current wireless channel, T will not be able to imitate A's fingerprint by using A's FIR filter.

\section{\texorpdfstring{D\MakeLowercase{eep}R\MakeLowercase{adio}ID}{DeepRadioID} FIR Optimization}\label{sec:filt_design}

In this section, we describe in details the \emph{DeepRadioID} FIR Filter Optimization module. We first provide some background notions in Section \ref{sec:background}, followed by our FIR-based waveform modification approach in Section \ref{sec:firbasedmod}. We then introduce the Waveform Optimization Problem (WOP) in Section \ref{sec:gradient}

\subsection{Background Notions and Definitions}\label{sec:background}
Let us define as \emph{input} a set of $N$ consecutive I/Q samples that constitute an input to the classifier. Let us also define as \emph{slice} a set of $S$ inputs, and as batch a set of $B$ slices. Let us label the $D$ devices being classified with a label between 1 and $D$. We model the classifier as as a function $f : \mc X \rightarrow \mc Y$, where $\mc X \subseteq \mathbb{C}^{N}$ and $\mc Y \subseteq \mathbb{R}^{D}$ represent respectively the spaces of the classifier's input (\textit{i.e.}, an example) and output (\textit{i.e.}, a probability distribution over the set of $D$ devices).
Specifically, the output of the classifier can be represented as a vector $(f_1, f_2, ..., f_D)\in\mc Y$, where the $i$-th component denote the probability that the input fed to the CNN belongs to device $i$.

\textit{DeepRadioID} relies on discrete causal finite impulse response filters (in short, FIRs) to achieve real-time adaptive waveform modification. FIRs present several advantages -- first, causal filters do not depend on future inputs, but only on past and present ones. Second, they can be represented as a weighted and finite term sum, which allows to \textit{accurately predict the output of the FIR for any given input}. More formally, a FIR is described by a finite sequence $\bs \phi$ of $M$ \textit{filter taps}, i.e., $\bs \phi=(\phi_1,\phi_2,\dots,\phi_M)$. 
For any input $\mf x \in \mc X$, the filtered $n$-th element $\hat{x}[n] \in \hat{\mf x}$ can be written as  
\begin{equation} \label{eq:FIR:general}
    \hat{x}[n] = \sum_{j=0}^{M-1} \phi_j x[n-j]
\end{equation}

Since both wireless channel and hardware impairments operate in the complex domain by rotating and amplifying/attenuating the amplitude of the signal, \textit{we can manipulate the position in the complex plane of the transmitted I/Q symbols}. By using complex-valued filter taps, \textit{i.e.}, $\phi_k\in \mb C$ for all $k=0,1,\dots,M-1$, we can rewrite Eq. \eqref{eq:FIR:general} as follows:

\begin{align} 
\hat{x}[n] & = \sum_{k=0}^{M-1} (\phi^R_k +j\phi^I_k) (x^R[n-k]+jx^I[n-k]) \nonumber \\
& = \hat{x}^R[n] +  j \hat{x}^I[n]\label{eq:FIR:complex}
\end{align}
\noindent
where $x^R_k[n] = \mbox{Re}\{x_k[n]\}$, $x^I_k[n] = \mbox{Im}\{x_k[n]\}$, $\phi^R_k = \mbox{Re}\{\phi_k\}$ and $\phi^I_k = \mbox{Im}\{\phi_k\}$. Eq. \eqref{eq:FIR:complex} clearly shows that it is possible to manipulate the input sequence by filtering each I/Q sample. For example, to rotate all I/Q samples by $\theta=\pi/4$ radiants and halve their amplitude, we may set $\phi_1 = \frac{1}{2} \exp^{j\frac{\pi}{4}}$ and $\phi_k = 0$ for all $k>1$. Similarly, other complex manipulations can be obtained by fine-tuning filter taps.

\subsection{FIR-based Waveform Modification}\label{sec:firbasedmod}


Although channel equalization can effectively reduce the effect of channel distortions on the position of the received I/Q samples, the algorithms involved are generally not perfect and only \textit{partially} counteract phase and amplitude variations caused by the channel. For this reason, we must devise techniques to dynamically adapt to rapidly changing channel conditions (\textit{e.g.}, fast-fading/multi-path) and thus improve the fingerprinting accuracy for a given device.

\begin{figure}[!h]
    \centering
    \includegraphics[width=\columnwidth]{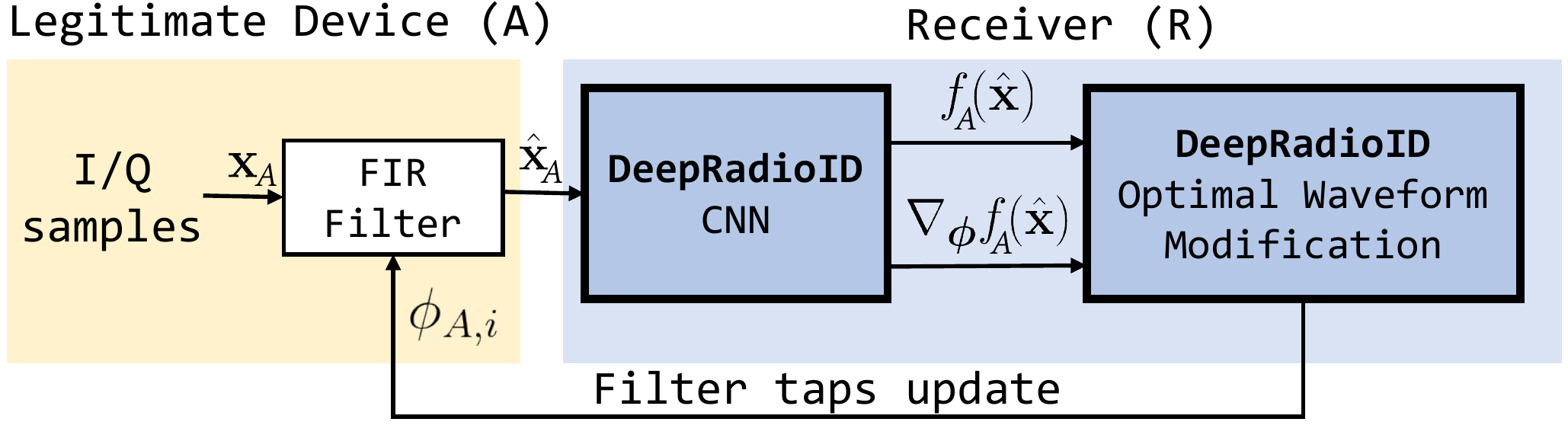}
    \caption{Waveform modification optimization loop.}
    \label{fig:DeepRadioID}
\end{figure}

\textit{DeepRadioID} leverages FIR filters to maximize the accuracy of the classifier by dynamically counterbalancing inaccurate channel equalization. Figure \ref{fig:DeepRadioID} shows a block diagram of the waveform modification optimization loop performed by \emph{DeepRadioID}. Specifically, we add a FIR filter before the first CNN layer. This additional layer uses FIRs to manipulate the input example according to Eq.~\eqref{eq:FIR:complex}. The corresponding output sequence is then fed to the CNN. 

As shown in Figure \ref{fig:arch}, let $A$ be the target device for which we want to improve the detection accuracy of the CNN, and let $\bs \phi_{A,i}$ be the filter taps associated to the target device at the $i$-th optimization step. By using the filtering-based waveform modification on the input sequence $\bf x$, the output $f_A(\hat{\bf x}) \in\mc Y$ of the classifier with respect to the filtered sequence $\hat{\bf x}$ can be written as a function of the filter taps ${\bs \phi}_{A,i}$. Specifically, we have that 
\begin{equation} \label{eq:output}
    f_A(\hat{\bf x}) = f_A(\bf x, {\bs \phi}_{A,i})
\end{equation}

Eq.~\eqref{eq:output} clearly shows that the accuracy of the classifier depends on the actual FIR tap vector ${\bs \phi}_{A,i}$. Thus, we are interested in devising mechanisms to optimally manipulate ${\bs \phi}_{A,i}$ such that (i) the classification accuracy for the target device is maximized (Section \ref{sec:gradient}); and (ii) the waveform modification does not negatively impact the BER of data transmission activities (Section \ref{sec:ber}). To simplify the notation, henceforth we will remove the $i$ subscript.

\subsection{Waveform Optimization Problem (WOP)} \label{sec:gradient}

We can now formally define the objective of \textit{DeepRadioID} as follows: (i) maximize the accuracy of the classifier for a specific target device $A$; and (ii) to guarantee that the resulting BER does not exceed a given maximum tolerable threshold $\mathrm{BER_{max}}$. Since we aim at achieving channel-resilient waveform modification, we need to compute a FIR parameter configuration ${\bs \phi}_A$ that can be used for multiple consecutive transmissions. It is worth mentioning that to compute different ${\bs \phi}_A$ values for each single input $\bf x$ is inefficient in many cases. Indeed, the obtained FIR would be effective for the considered input only, \textit{i.e.}, if applied to another input sequence $\bf x' \neq \bf x$, the FIR might decrease the accuracy of the classifier. Thus, maximizing the accuracy with respect to a single input $\bf x$ might result in poor performance.

To overcome the above problem, rather than maximizing the accuracy of the classifier on an input-by-input basis, we compute the the FIR ${\bs \phi}_A$ that maximizes the activation probability $f_A$ over a set of $S$ consecutive inputs, \textit{i.e.}, a slice.

The Waveform Optimization Problem (WOP) can be then defined as follows:
\begin{align}
\underset{\bs \phi}{\text{maximize}} & \hspace{0.2cm} \frac{1}{S} \sum_{s=1}^S f_A({\bf x}_s, \bs \phi) \label{prob:optimal:util} \tag{WOP} \\
    \text{subject to} & \hspace{0.1cm}  \mathrm{BER}_A(\mathrm{{\bf x}_s, \bs \phi)} \leq \mathrm{BER_{max}}, \hspace{0.2cm} \forall{s=1,2,\dots,S} \label{prob:optimal:constraint} \tag{C1}
\end{align}
\noindent
where the objective function represents the per-slice average activation probability for device $A$, ${\bf x}_s$ is the $s$-th input of the slice, and the $\mathrm{BER}_A(\cdot)$ represents the BER function corresponding to transmissions from target node $A$.

It is worth mentioning that the function $f_A(\mathbf{x}_s, \boldsymbol{\phi})$ represents the CNN, and thus it outputs the probability that the input I/Q samples $\mathbf{x}$ belong to device $A$. Thus, by solving the WOP, we compute a FIR that maximizes the activation probability of the neuron associated to $A$. 

Problem \eqref{prob:optimal:util} is significantly challenging because (i) the function $f_A$ is CNN-specific and depends from a very high number of parameters (generally in the order of millions), it is highly non-linear and to the best of our knowledge, there are no mathematical closed-form expressions for such a function, even for relatively small CNNs; (ii) the maximum BER constraint \eqref{prob:optimal:constraint} depends from numerous device-specific parameters (\textit{e.g.}, modulation, coding, transmission power and SNR) and it is generally non-linear.

Notwithstanding the above challenges, and as we will discuss in detail in Section \ref{sec:ber}, the impact of the waveform modification procedure on the BER of communications among the receiver and the target device $A$ is negligible. In fact, as shown in Figure \ref{fig:arch}, \textit{DeepRadioID} embeds a FIR Filter Compensation module that uses peculiar features of FIR filters, \textit{e.g.}, their Fourier transform, to successfully reconstruct the original transmitted unfiltered sequence of I/Q symbols. This compensation procedure effectively removes any coupling between waveform modification procedures and BER, \textit{i.e.}, $\mathrm{BER}_A(\mathrm{\bf x, \bs \phi)} \approx \mathrm{BER}_A(\mathrm{\bf x})$.
Accordingly, it is possible to relax Constraint \eqref{prob:optimal:constraint} by removing it from the optimization problem \eqref{prob:optimal:util}.

The relaxed WOP can be formulated as  
\begin{align}
\underset{\bs \phi}{\text{maximize}} & \hspace{0.2cm} \sum_{s=1}^S f_A({\bf x}_s, \bs \phi) \label{prob:optimal:util:2} \tag{RWOP}
\end{align}
\noindent
where we have also omitted the constant term $1/S$.

\subsubsection{Solving the RWOP}

As already mentioned, $f_A$ is non-linear and generally does not possess any useful property in terms of monotonicity, concavity and existence of a global maximizer. 
However, for any input $\bf x$ of the slice, by using \textit{back-propagation} and the chain rule of derivatives it is possible to let the CNN compute the gradient $\nabla_{\hat{\bf x}}f_A(\hat{\bf x})$ of the classification function $f_A$ with respect to the filtered input sequence $\hat{\bf x}$. It is worth noting that $\nabla_{\hat{\bf x}}f_A(\hat{\bf x})$ shows how different input sequences affect the accuracy of the classification function. Nevertheless, we are interested in evaluating the gradients $\nabla_{\bs \phi}f(\hat{\bf x})$ to predict how the accuracy of the classifier varies as a function of the FIR filtering function.
Hence, we need to extend back-propagation to the waveform modification block. 

From Eq. \eqref{eq:FIR:complex}, $\hat{\bf x}$ is a function of $\bs \phi$, thus the gradient of $f_A$ with respect to the filter taps $\bs \phi$ can be computed as
\begin{equation} \label{eq:chain:vector}
    \nabla_{\bs \phi}f_A(\hat{\bf x}) = J_{f_A}(\bs \phi)^\top \cdot \nabla_{\hat{\bf x}}f_A(\hat{\bf x})
\end{equation}
\noindent
where $J_{f_A}(\bs \phi)$ is the Jacobian matrix of $f_A(\bf x, \bs \phi)$ with respect to $\bs \phi$, $\top$ is the transposition operator, and $\cdot$ stands for matrix dot product. 

From Eq. \eqref{eq:chain:vector} and Eq. \eqref{eq:FIR:complex}, each element in $\nabla_{\bs \phi}f_A(\hat{\bf x})$ can be written as
\begin{align} \label{eq:chain:partial}
    \frac{\partial f_A(\bf x, \bs \phi) }{\partial \phi^Z_k} & = \sum_{n = 1}^N \left( \frac{\partial f_A(\bf x, \bs \phi)}{\partial \hat{x}^R[n]} \frac{\partial \hat{x}^R[n] }{\partial \phi^Z_k} + \frac{\partial f_A(\bf x, \bs \phi)}{\partial \hat{x}^I[n]} \frac{\partial \hat{x}^I[n] }{\partial \phi^Z_k}\right)
\end{align}
\noindent
where $k=0,1,\dots,M-1$, $N$ is the length of the input sequence and $Z\in\{R,I\}$.

By using Eq. \eqref{eq:FIR:complex}, $\frac{\partial \hat{x}^R[n] }{\partial \phi^Z_k}$ and $\frac{\partial \hat{x}^I[n] }{\partial \phi^Z_k}$ in Eq. \eqref{eq:chain:partial} are computed as follows:
\begin{align}
    \frac{\partial \hat{x}^R[n] }{\partial \phi^R_k} & = \frac{\partial \hat{x}^I[n] }{\partial \phi^I_k} = x^R[M-1+n-k] \label{eq:chain:partial:1} \\
    \frac{\partial \hat{x}^I[n] }{\partial \phi^R_k} & = - \frac{\partial \hat{x}^R[n] }{\partial \phi^I_k} = x^I[M-1+n-k] \label{eq:chain:partial:2}
\end{align}

The above analysis shows that the relationship between the waveform modification and classification processes can be described by a set of gradients. Most importantly, they can be used to devise effective optimization algorithms that solve Problem \eqref{prob:optimal:util:2}.

In Section \ref{sec:ncg}, we design an algorithm to solve Problem \eqref{prob:optimal:util:2} and compute the optimal FIR filter parameters $\bs \phi$ by using the Nonlinear Conjugate Gradient (NCG) method and the gradients computed in Eq. \eqref{eq:chain:partial:1} and  Eq. \eqref{eq:chain:partial:2}. While our simulation results have shown that NCG is more accurate than other gradient-based optimization algorithms (\textit{e.g.}, gradient descent algorithms), we remark that \textit{DeepRadioID} is independent of the actual algorithm used to compute $\bs \phi$, and other approaches can be used to solve Problem \eqref{prob:optimal:util:2}.

\subsubsection{Filter taps computation through NCG} \label{sec:ncg}

As shown in Figure \ref{fig:arch} and discussed in Section \ref{sec:system}, \textit{DeepRadioID} iteratively adapts to channel fluctuations by periodically updating the filter taps associated to any given target device $A$. 
For the sake of generality, we refer to this periodic update as an \textit{optimization epoch}, and a new epoch is started as soon as one or more \textit{triggering events} are detected by \textit{DeepRadioID}. Triggering events can be either cyclic, \textit{e.g.}, timer timeout, or occasional, \textit{e.g.}, the accuracy for a target devices falls below a minimum desired threshold.   

For each epoch $i$, let $t=1,2,\dots,T$ denote the iteration counter of the optimization algorithm. At each iteration $t$ of the algorithm, the filter taps are updated according to the following iterative rule
\begin{equation} \label{eq:CGM:general}
    {\bs \phi}^{(t)} = {\bs \phi}^{(t-1)} + {\bs \alpha^{(t)}} {\bf p^{(t)}}
\end{equation}

In Eq. \eqref{eq:CGM:general}, ${\bf p^{(t)}}$ and ${\bs \alpha^{(t)}}$ represent the search direction and update step of the algorithm, respectively. To put it simple, ${\bf p^{(t)}}$ gives us information on the direction to be explored, while ${\bs \alpha^{(t)}}$ tells us how large the exploration step taken in that direction should be. More in detail, the two terms are computed as follows:
\begin{align}
    {\bf p^{(t)}} &= \sum_{s=1}^S \left( \nabla_{{\bs \phi}} f_A({\bf x}_s,{\bs \phi}^{(t-1)}) \right) + {\bs \beta}^{(t)} {\bf p^{(t-1)}} \label{eq:CGM:direction} \\
    {\bs \alpha}^{(t)} & = \argmax_{\bs \alpha} \sum_{s=1}^S f_A({\bf x}_s,{\bs \phi}^{(t-1)} + {\bs \alpha} {\bf p^{(t)}}) \label{eq:CGM:step}
\end{align}
\noindent
where gradients derive from Eq. \eqref{eq:chain:partial:1} and Eq. \eqref{eq:chain:partial:2}, $\beta^{(1)} = 0$ and $p^{(0)} = 0$.
The parameter ${\bs \beta}^{(t)}$ is defined as 
\begin{equation}
        {\bs \beta}^{(t)} = \frac{|| \sum_{s=1}^S \left( \nabla_{{\bs \phi}} f_A({\bf x}_s,{\bs \phi}^{(t-1)}) \right) ||^2_2}{|| \sum_{s=1}^S \left( \nabla_{{\bs \phi}} f_A({\bf x}_s,{\bs \phi}^{(t-2)}) \right) ||^2_2} \label{eq:CGM:beta}
\end{equation}
\noindent
and is generally referred to as the \textit{conjugate gradient} (update) parameter used in NCG methods to improve the space exploration process by speeding up the convergence of the algorithm \cite{hestenes1952methods}. 

Interestingly enough, ${\bf p^{(1)}} = \sum_{s=1}^S \left( \nabla_{{\bs \phi}} f_A({\bf x}_s,{\bs \phi}^{(0)}) \right)$ when $t=1$. That is, the first iteration of the NCG algorithm corresponds to a classic gradient descent. Also, ${\bs \alpha}^{(t)}$ in Eq. \eqref{eq:CGM:step} is computed through line-search algorithms. While both exact and approximated line search algorithms can be considered, there are few aspects that need to be considered when implementing Eq. \eqref{eq:CGM:step}. Indeed, since the function $f_A$ is highly non-linear and has no closed-form representation, to compute Eq. \eqref{eq:CGM:step} requires the continuous evaluation of $\sum_{s=1}^S f_A({\bf x}_s,{\bs \phi}^{(t-1)} + {\bs \alpha} {\bf p^{(t)}})$ and its first and second order derivatives. For this reason, in some cases it might be computationally expensive to run exact line search algorithms on $f_A$, and approximated line search algorithms are to be preferred. For example, to speed-up the computation of ${\bs \alpha}^{(t)}$, we can consider a secant method approximation where the second derivatives of $f_A$ are approximated by using the first order derivatives computed in Eq. \eqref{eq:chain:partial:1} and Eq. \eqref{eq:chain:partial:2}.

\section{\texorpdfstring{D\MakeLowercase{eep}R\MakeLowercase{adio}ID}{DeepRadioID} FIR Compensation}\label{sec:ber}

Although the waveform filtering process is beneficial to the classification process as we can optimally modify the waveform generated by a given target device, it may negatively affect the quality of transmitted data. Indeed, moving I/Q symbols within the complex plane might impact the demodulation and decoding process, thus increasing the BER associated to the received waveform. 
Furthermore, the expression of the BER in Constraint \eqref{prob:optimal:constraint} is non-linear and possesses exact and/or approximated closed-form representations only in a limited number of cases (\textit{e.g.}, fading channels with known distributions and low-order modulations).

The above discussion shows that to tackle the BER constraint in Problem \eqref{prob:optimal:util} is challenging, since we need to devise mechanisms that are generic enough to be used with any modulation, coding and channel distributions. To overcome the above issues, we observe that our waveform modification relies on Eq.~\eqref{eq:FIR:general}, which clearly represents a discrete convolution between the input sequence $\bf x$ and the filter taps $\bs \phi$. For the sake of illustration, in the following we use the familiar model
\begin{equation}
    z[n] = ({\bf h} \circledast {\bf \hat{x}})[n] + w[n]
\end{equation}
\noindent
where each received I/Q symbol $z[n]$ is written as the sum of a noise term $w[n]$ (typically Additive white Gaussian noise) and the $n$-th element of the discrete convolution ${\bf h} \circledast {\bf \hat{x}}$ between the channel $h$ and the transmitted filtered sequence ${\bf \hat{x}}$. 
From Eq.~\eqref{eq:FIR:general}, we also have that $\hat{{\bf x}} = {\bf x} \circledast \phi$. 
Thus, the discrete Fourier transform (DFT) of ${\bf z}$ and ${\bf \hat{x}}$ can be written as follows:\vspace{-0.2cm}

\begin{align} \label{eq:DFT}
    Z(\omega) & = H(\omega) \hat{X}(\omega) + W(\omega) \nonumber \\
         & = H(\omega) X(\omega) \Phi(\omega) + W(\omega)
\end{align}
\begin{equation} \label{eq:DFT:X}
    X(\omega) = \frac{Z(\omega) - W(\omega)}{H(\omega) \Phi(\omega)}
\end{equation}
\noindent
where we have used capital letters to indicate DFTs.

Eq.~\eqref{eq:DFT:X} shows that to reconstruct the original unfiltered I/Q sequence is possible by computing the inverse DFT of each component of the received signal. Furthermore, FIR compensation implies $\mathrm{BER}_A(\mathrm{\bf x, \bs \phi)} = \mathrm{BER}_A(\mathrm{\bf x})$. Since, the optimization variable of Problem \eqref{prob:optimal:util} is $\bs \phi$, Constraint \eqref{prob:optimal:constraint} does not depend on $\bs \phi$ anymore, and thus it can be removed from Problem \eqref{prob:optimal:util}. Finally, note that the above FIR compensation method is independent of the underlying modulation and coding scheme. It means, that FIR compensation is a general approach, and it can be successfully used to tackle Constraint \eqref{prob:optimal:constraint}.

Despite the above properties, one might argue that in general it is not possible to compute a perfect estimation of $W(\omega)$ and $H(\omega)$. However, modern wireless networks embed estimation mechanisms that are almost always able to compute fairly accurate estimations $\tilde{H}(\omega)$ and $\tilde{N}(\omega)$, \textit{e.g.}, through training sequences and pilots \cite{sourour2004frequency}, and the effect of the FIR filter can thus be compensated to a significant extent. To validate this crucial assumption, we ran a number of experiments on our experimental testbed to evaluate the impact of \emph{DeepRadioID}'s FIR filtering on the packet error rate (PER) and throughput ($\theta$) of a wireless transmission. 

\begin{figure}[!h]
    \centering
    \includegraphics[width=\columnwidth]{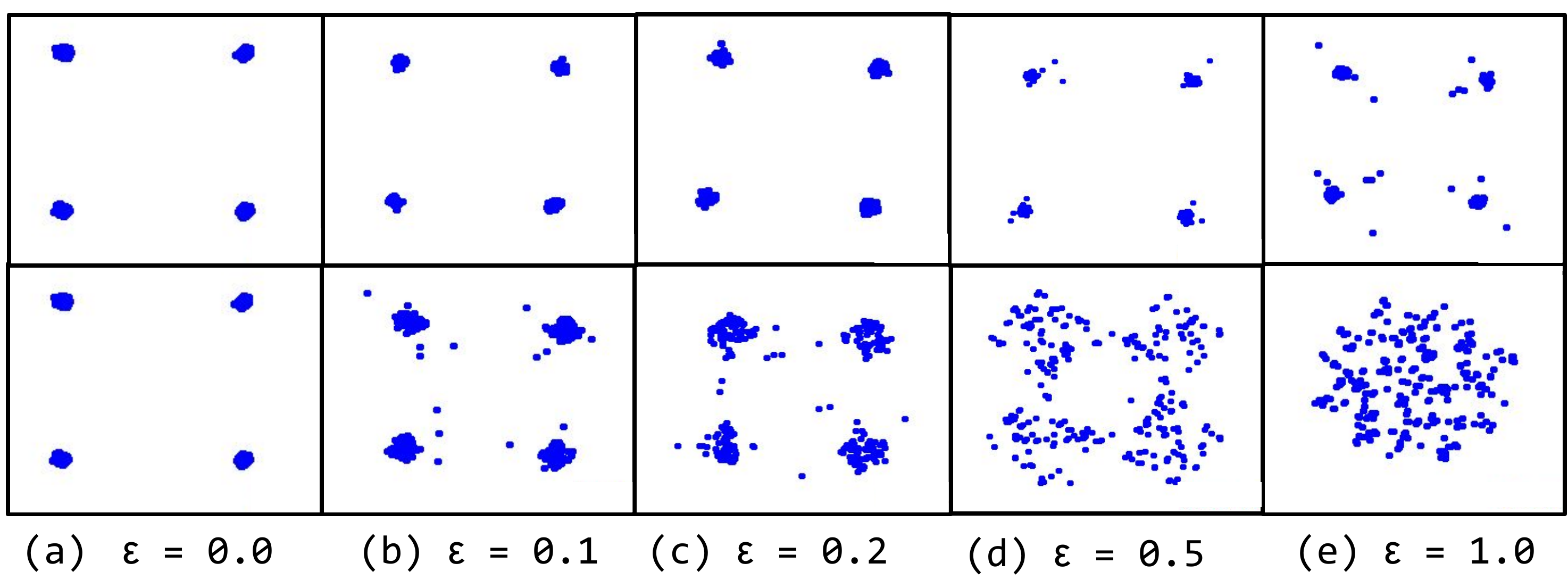}
    \caption{The effect of FIR filtering at the receiver's side for different values of $\epsilon$. Top and bottom sides show respectively the received constellations with/without FIR compensation.\vspace{-0.5cm}}
    \label{fig:FIR_effect}
\end{figure}

\begin{figure}[!h]
    \centering
    \includegraphics[width=0.5\columnwidth,angle=-90]{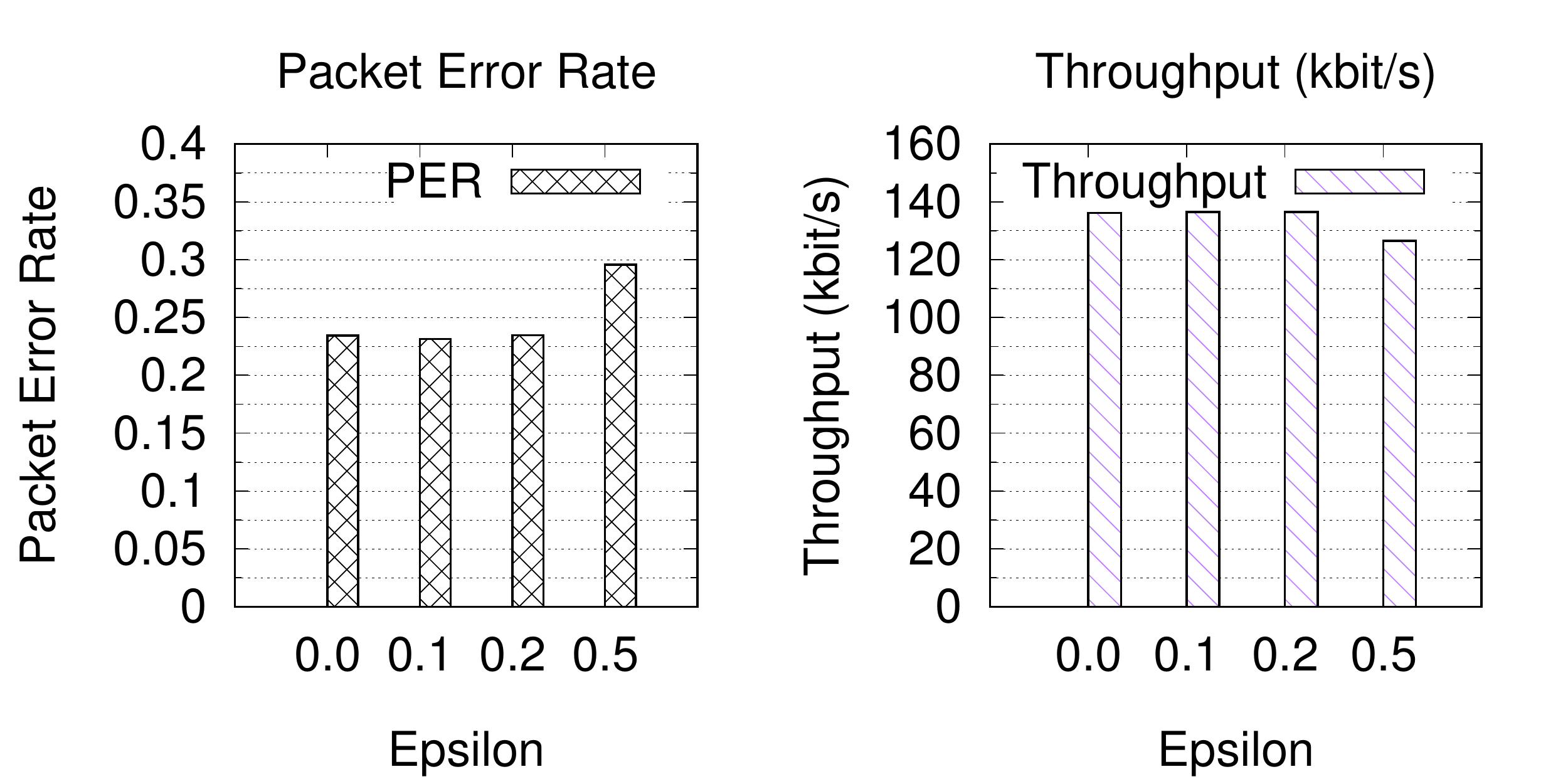}
    \caption{Packet error rate (PER) and Throughput ($\theta$) as a function of $\epsilon$.}
    \label{fig:PER_Throughput}
\end{figure}

Figure \ref{fig:FIR_effect} shows the received constellations of a QPSK-modulated WiFi transmission where the payload I/Q samples are multiplied in the frequency domain (\textit{i.e.}, FIR filtering) with a random I/Q tap with $ I \in [1 -\epsilon, 1 + \epsilon]$ and $Q \in [0 -\epsilon, 0 + \epsilon])$. The $\epsilon$ parameter represents the relative magnitude of the filter with respect to no filtering, \textit{i.e.}, $\epsilon = 0$. The filtering is then compensated at the receiver's side by using Eq.~\eqref{eq:DFT:X}. 

Due to the imperfection in channel compensation, we notice than some noise is indeed introduced by the FIR filtering irrespective of our compensation. However, these imperfections do not translate in a significant PER increase. Figure \ref{fig:PER_Throughput} shows the PER and $\theta$ as a function of $\epsilon$, which respectively increase and decrease of about 6\% and 0.5~kbit/s in the worst case of $\epsilon = 0.5$. However, according to our experiments in Section \ref{sec:exp_res}, the $\epsilon$ value is typically below 0.2, meaning a PER increase <1\% and a $\theta$ loss <0.2 kbit/s (0.2\%). 


\section{Experimental Results}\label{sec:exp_res}

In this section, we report the results obtained through extensive experimentation on a practical software-defined radio testbed (Section \ref{sec:testbed}), as well as on three datasets of WiFi and ADS-B transmissions obtained through the DARPA RFMLS program (Section \ref{sec:dataset_res}). 

\subsection{Radio Testbed Setup}\label{sec:testbed}



Our experimental testbed is composed by twenty software-defined USRP radios acting as transmitters and one USRP acting as receiver. Each USRP has been equipped with a CBX 1200-6000 MHz daughterboard with 40 MHz instantaneous bandwidth \cite{CBX} and one VERT2450 antennas \cite{Vert2450}. Therefore, the RF components of each USRP are nominally-identical. Furthermore, each USRP device sends the same baseband signal , \textit{i.e.}, an IEEE 802.11a/g (WiFi) frame repeated over and over again, to make sure that the deep learning model is learning the hardware impairments and not data patterns. 

\begin{figure}[!h]
    \centering
    \includegraphics[width=\columnwidth]{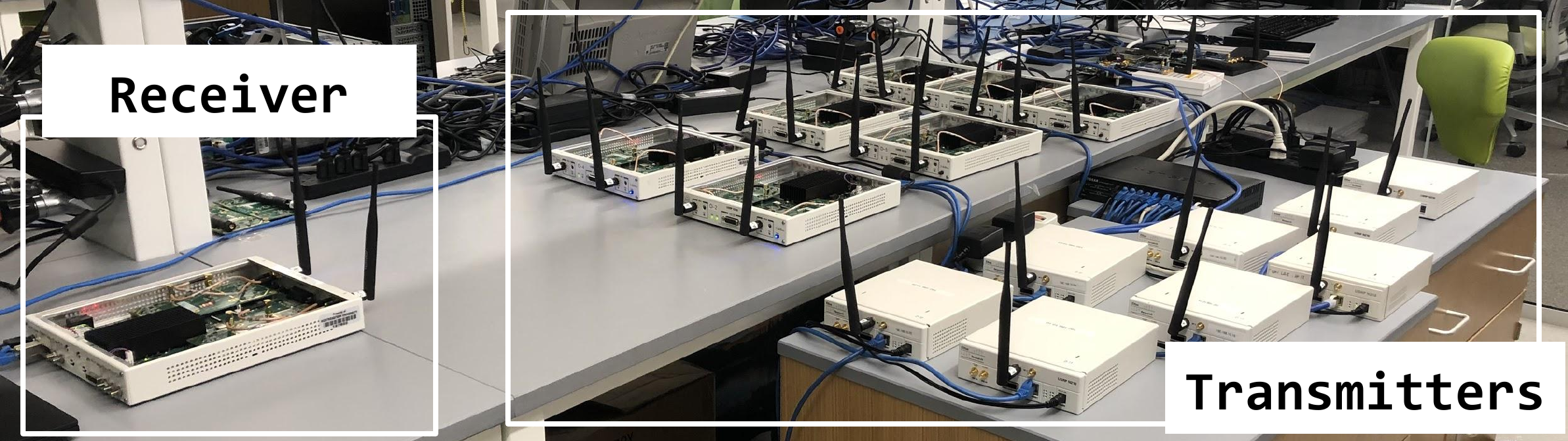}
    \caption{DeepRadioID Experimental Testbed.}
    \label{fig:testbed}
\end{figure}

The baseband signal is generated through Gnuradio and then streamed to the selected SDR for over-the-air wireless transmission. The receiver SDR samples the incoming signals at $10\ \textrm{MS/s}$ sampling rate at center frequency of $2.432\  \textrm{GHz}$. The collected baseband signal is then channel-equalized using IEEE 802.11 pilots and training sequences \cite{sourour2004frequency}. Next, the payload I/Q samples are extracted and partitioned into a \textit{sample}. In our experiments, we fix the sample length to $ 48 \cdot 6 = 288$ I/Q values, corresponding to 6 OFDM symbols containing 48 payload I/Q values. Each of these samples are then used for training and classification. 

\subsection{Deep Learning Architecture}\label{sec:deep_learn_arch}

\begin{figure}[!h]
    \centering
    \includegraphics[width=\columnwidth]{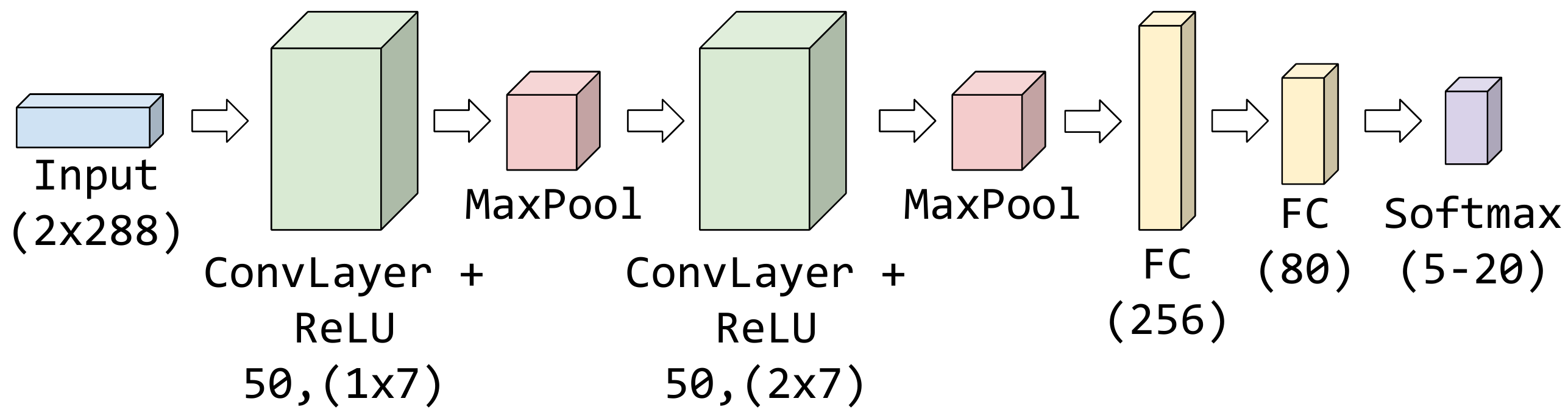}
    \caption{CNN used in our radio testbed experiments.\vspace{-0.3cm}}
    \label{fig:cnn_testbed}
\end{figure}

We use the CNN architecture reported in \cite{5_8466371} and depicted in Figure \ref{fig:cnn_testbed}. Specifically, each I/Q input sequence is represented as a two-dimensional real-valued tensor of size $2 \times 288$. This is then fed to the first convolutional  layer (ConvLayer), which consists of 50 filters  each of size $1 \times 7$. Each filter learns a 7-sample variation in time over the I or Q dimension separately, to generate 50 distinct feature maps over the complete input sample. Similarly, the second ConvLayer has 50 filters each of size $2 \times 7$.  Each ConvLayer is followed by a Rectified Linear Unit (ReLU) activation and a maximum pooling (MaxPool) layer with filters of size $2 \times 2$ and stride $1$, to perform a pre-determined non-linear transformation on each element of the convolved output.

The output of the second convolution layer is then provided as input to the first fully connected layer, which has 256 neurons. A second fully connected layer of 80 neurons is added to extract higher level non-linear combinations of the features extracted from previous layers, which are finally passed to a classifier layer. To overcome overfitting, we set the dropout rate to 50\% at the dense layers. A \textit{softmax} classifier is used in the last layer to output the probabilities of each sample being fed to the CNN.  

We use an $\ell_2$ regularization parameter $\lambda=0.0001$. The weights of the network are trained using Adam optimizer with a learning rate of $l = 0.0001$. We minimize the prediction error through back-propagation, using categorical cross-entropy as a loss function computed on the classifier output. We implement our CNN architecture in Keras running on top of TensorFlow on a system with 8 NVIDIA Cuda enabled Tesla K80m GPU. To train our CNN, we constructed a dataset by performing 10 different transmissions of length 5 minutes for each of the 20 devices, each transmission spaced approximately 5 minutes in time. 

\subsubsection{Performance Metrics}\label{sec:train_test}

Hereafter, we will use the following two metrics to assess the performance of our learning system:

\begin{enumerate}
    \item \emph{Per-Slice Accuracy (PSA).} As in Section \ref{sec:gradient}, a \textit{slice} is a set of $S$ consecutive input. The PSA is thus defined as the average CNN fingerprinting accuracy on the $S$-sample slice. Since the FIR filter optimized by \emph{DeepRadioID} is computed on one slice, the PSA measures how much \emph{DeepRadioID} is able to increase the short-term accuracy of the CNN.
    \item \emph{Per-Batch Accuracy (PBA).} We define as \textit{batch} as a set of $B$ consecutive slices. The PBA is thus defined as the average CNN fingerprinting accuracy on the $B$-slice batch. The PBA measures the impact of the optimal FIR filter on the long-term accuracy of the CNN.
\end{enumerate}

In the following experiments, we use live-collected data (\textit{i.e.}, not coming from the training dataset) to compute the PSA and PBA. This data was collected 7 days after the data used for training the dataset was collected. To allow experiments' repeatability, we record a transmission from a given device for about one minute. Then, we select the first slice from the recording and compute the PSA with no FIR optimization. Next, we perform FIR optimization on the slice, re-filter each sample in the slice, and compute the PSA after FIR filtering. To compute the PBA, we apply the same FIR filter to the  $B \cdot N -1$ slices collected after the first one and then compute the average PSA on the $B$-slice batch.

\subsection{DeepRadioID Testbed Results}\label{sec:opt_res}

\begin{figure}[!h]
    \centering
    \includegraphics[width=0.5\columnwidth,angle=-90]{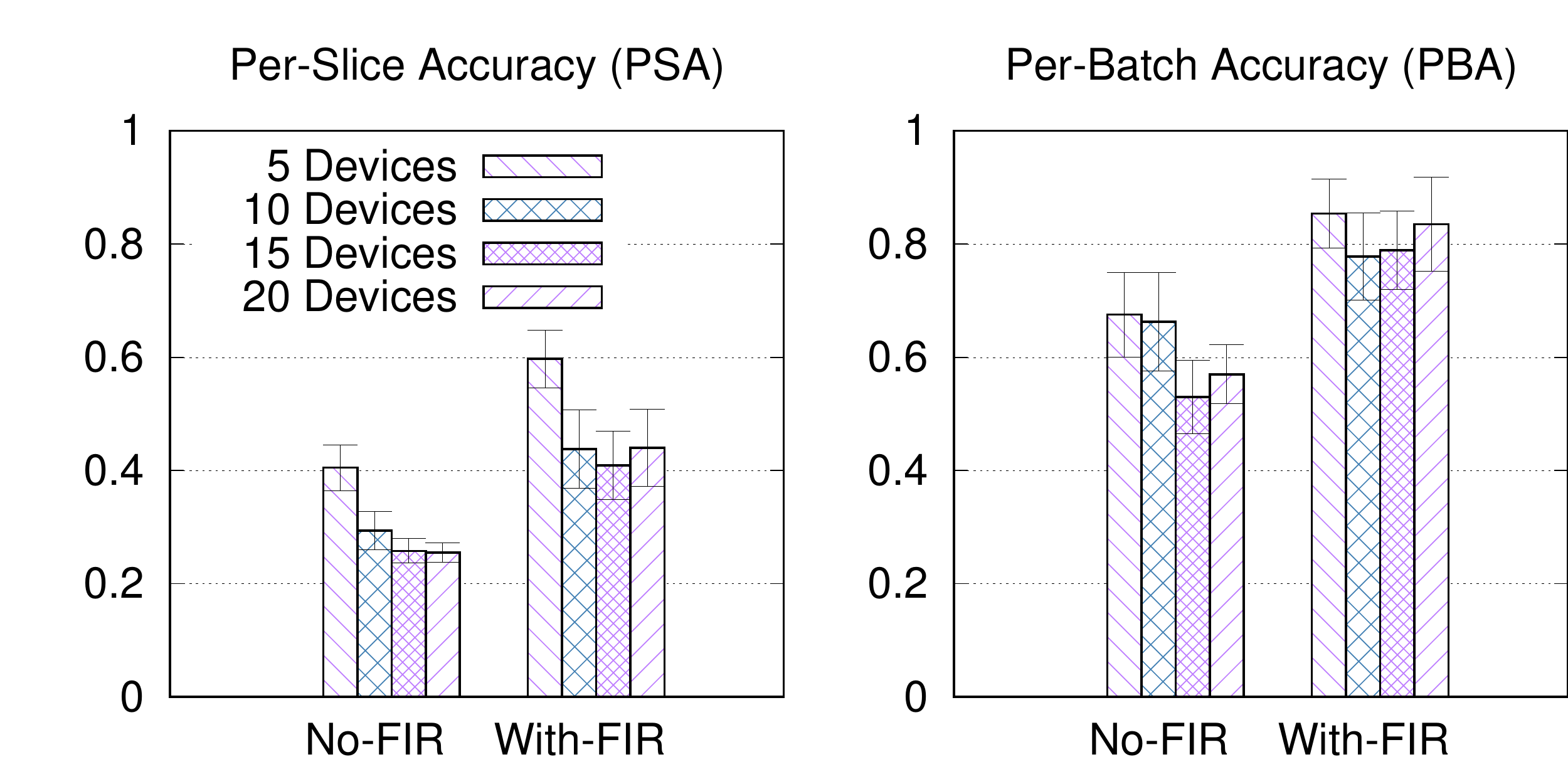}
    \caption{PSA and PBA, Experimental Testbed.\vspace{-0.3cm}}
    \label{fig:pba_testbed}
\end{figure}

Figure \ref{fig:pba_testbed} shows  the average PSA and PBA obtained by optimizing three devices over ten different recordings PSA as a function of the model size (\textit{i.e.}, 5 to 20 devices). We also show 95\% confidence intervals.

The results obtained in Figure \ref{fig:pba_testbed} show that \textit{DeepRadioID} is significantly effective in improving the PSA and PBA of our CNN-based fingerprinting system, which are improved by an average of about 57\% and 35\%, respectively. By accounting that the metrics are computed on live-collected data and on a testbed of nominally-identically radios, we consider these results remarkable. 

\begin{figure}[!h]
    \centering
    \includegraphics[width=0.42\columnwidth,angle=-90]{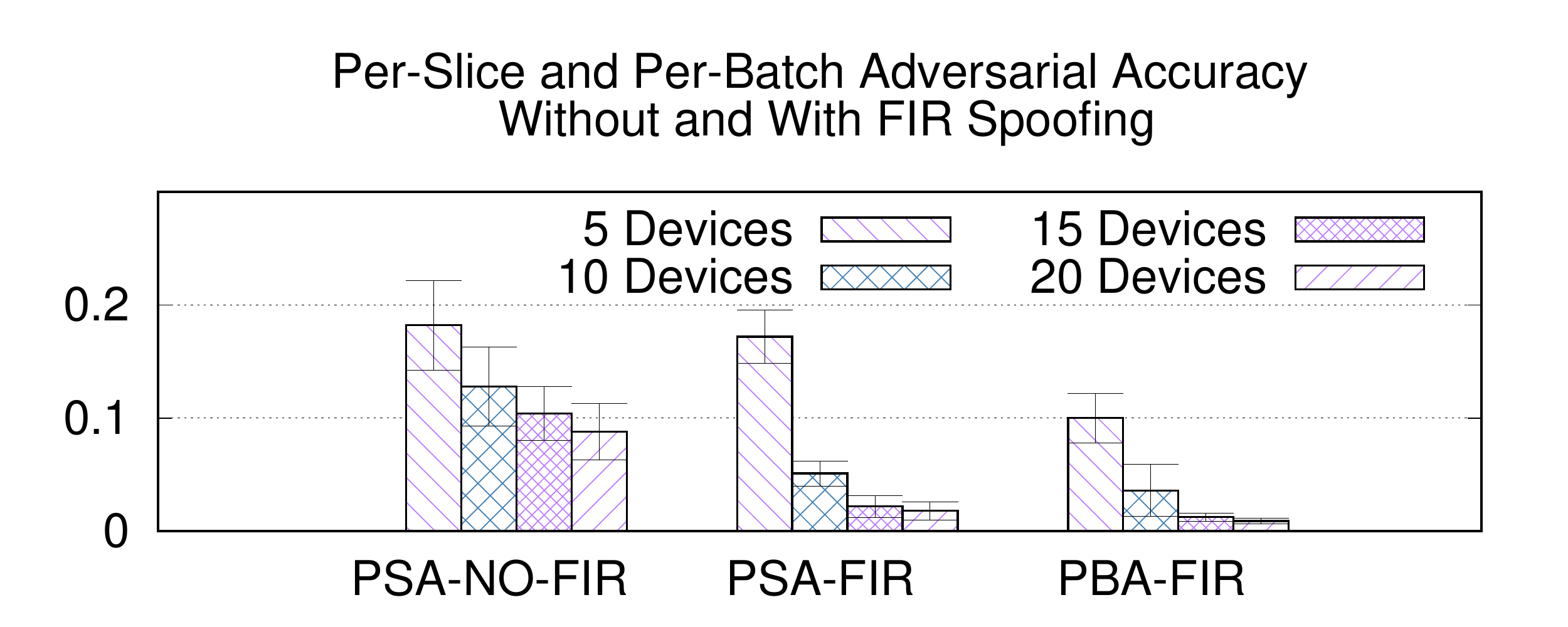}
    \caption{Adversarial Action, Experimental Testbed.\vspace{-0.3cm}}
    \label{fig:adv_testbed}
\end{figure}

Figure \ref{fig:adv_testbed} shows the PSA and the PBA obtained by an adversary trying to imitate another device's fingerprint by applying the same FIR. In these experiments, we fixed one device as adversary and used the FIR from other three devices to compute its PSA and PBA for each of its ten recordings. Figure \ref{fig:adv_testbed} shows that by using the legitimate device's FIR, the adversary transitions from an average PSA of 12\% to an average PSA of about 6\% (50\% decrease), corresponding to an average PBA of about 4\%. This ultimately confirms that since a FIR is optimized for a device's specific channel and impairments, an adversary cannot use a legitimate device's FIR to imitate its fingerprint.

\subsection{DeepRadioID Dataset Results}\label{sec:dataset_res}

To investigate \emph{DeepRadioID}'s performance on large-scale wireless systems and with different wireless standards, we consider (i) a 500-device dataset of IEEE 802.11a/g (WiFi) transmissions;  and (ii) a 500-airplane dataset of Automatic Dependent Surveillance -- Broadcast (ADS-B) beacons, both obtained through the DARPA RFMLS program. ADS-B is a surveillance transmission where an aircraft determines its position via satellite navigation and periodically broadcasts it at center frequency 1.090~GHz with sampling rate 1~MSPS and pulse position modulation. This makes ADS-B ideal to evaluate \emph{DeepRadioID}'s performance on different channel/modulation scenarios. For both the WiFi and ADS-B datasets, data collection was performed ``in the wild'' (\textit{i.e.}, no controlled environment) with a Tektronix RSA operating at 200 MSPS. For the WiFi dataset, as in Section \ref{sec:opt_res} we demodulated the transmissions and trained our models on the derived I/Q samples. To demonstrate the generality of \emph{DeepRadioID}, the ADSB model was instead trained on the unprocessed I/Q samples.

\begin{figure}[!h]
    \centering
    \includegraphics[width=\columnwidth]{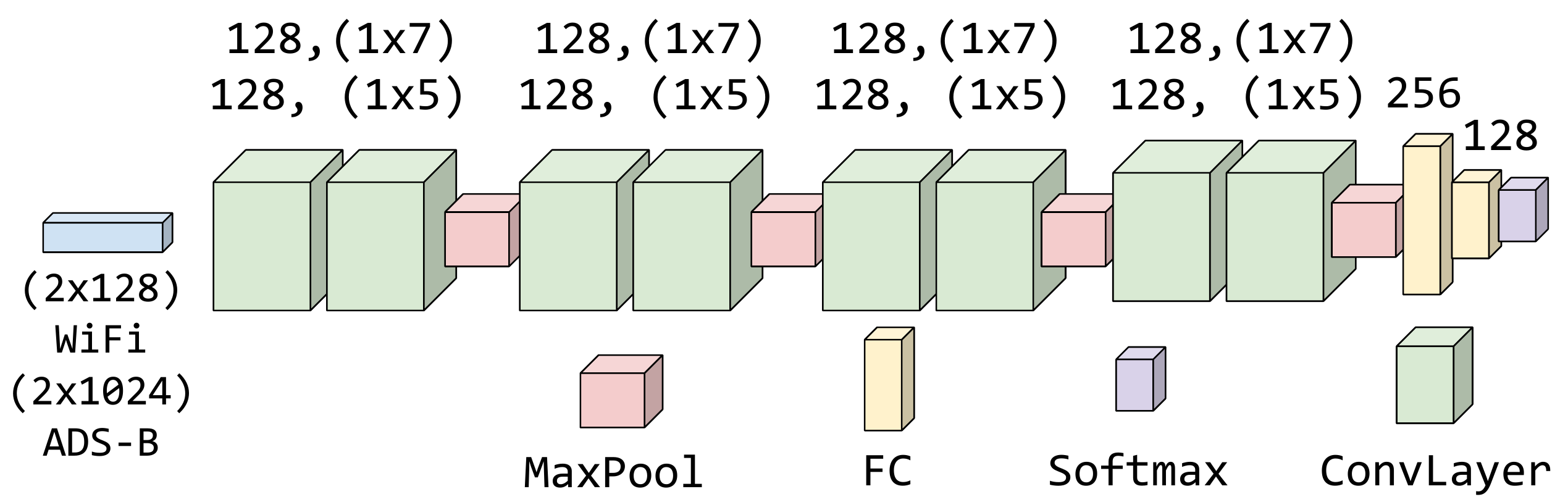}
    \caption{CNN used in our dataset experiments.\vspace{-0.3cm}}
    \label{fig:cnn_datasets}
\end{figure}

To handle the increased number of devices and experiment with a different CNN, we use the CNN architecture shown in Figure \ref{fig:cnn_datasets}. In the case of ADS-B we train our CNN on examples containing 1024 consecutive unprocessed I/Q samples. For WiFi, we only use 128 samples. Unless stated otherwise, PSA and PBA are computed on batches of 12 slices, each containing 25 inputs.

\begin{figure}[!h]
    \centering
    \includegraphics[width=0.5\columnwidth,angle=-90]{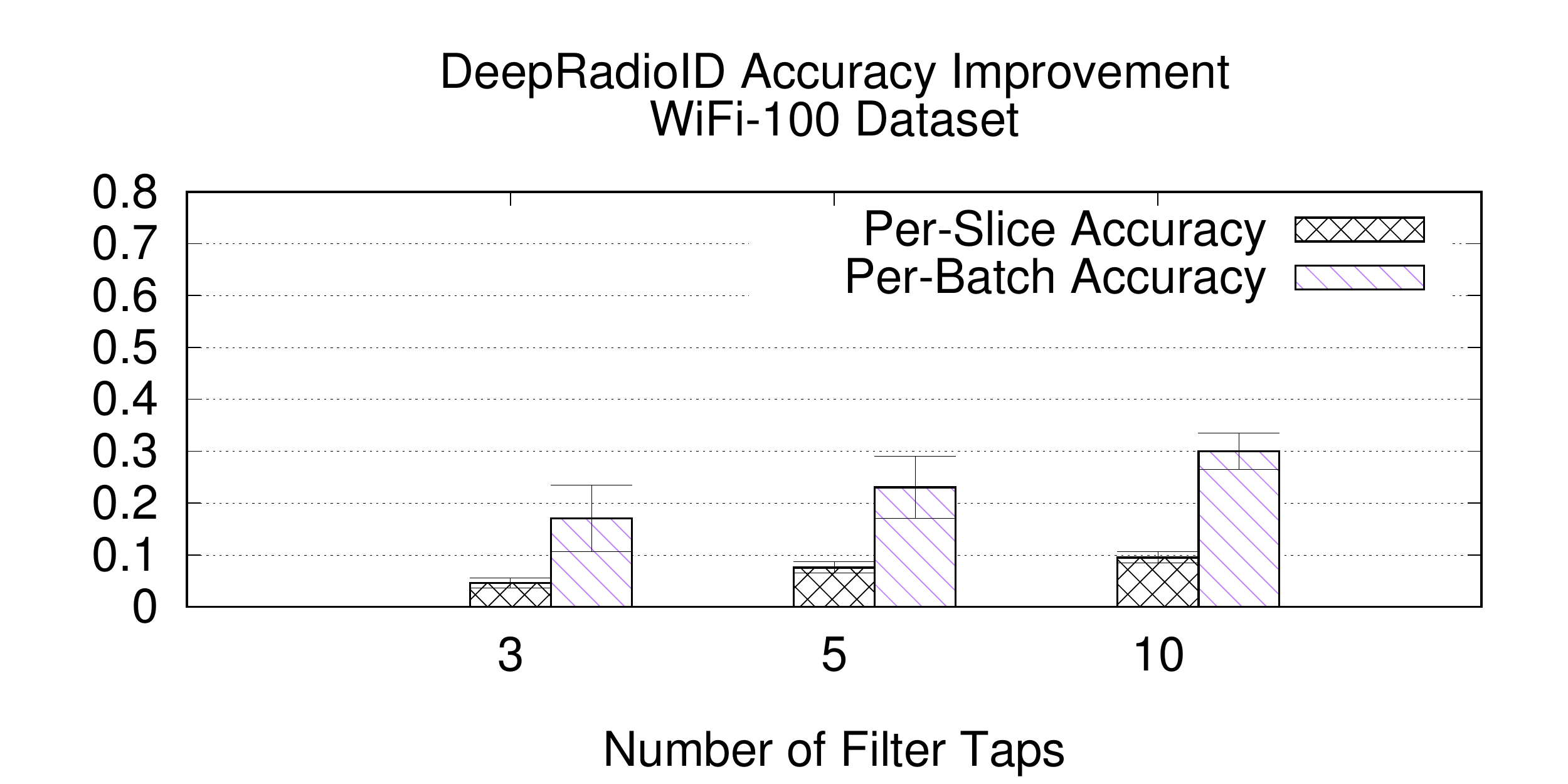}
    \caption{PSA/PBA Improvement, WiFi-100.\vspace{-0.2cm}}
    \label{fig:wifi100improvement}
\end{figure}

To compare our approach with the state of the art by Vo \emph{et al.}~\cite{12_vo2016fingerprinting}, which reports results on 93 devices, we trained our model on a subset of 100 devices, achieving PSA and PBA of .32 and .44, respectively. Figure \ref{fig:wifi100improvement} shows the PSA and PBA improvement brought by \textit{DeepRadioID} FIR filtering as function of the number of FIR taps. In particular, the PBA improvement is around 30\% when 10 FIR taps are used, which brings the accuracy to about 74\% on the average, outperforming \cite{12_vo2016fingerprinting} which is 47\%.

\begin{figure}[!h]
    \centering
    \includegraphics[width=0.50\columnwidth,angle=-90]{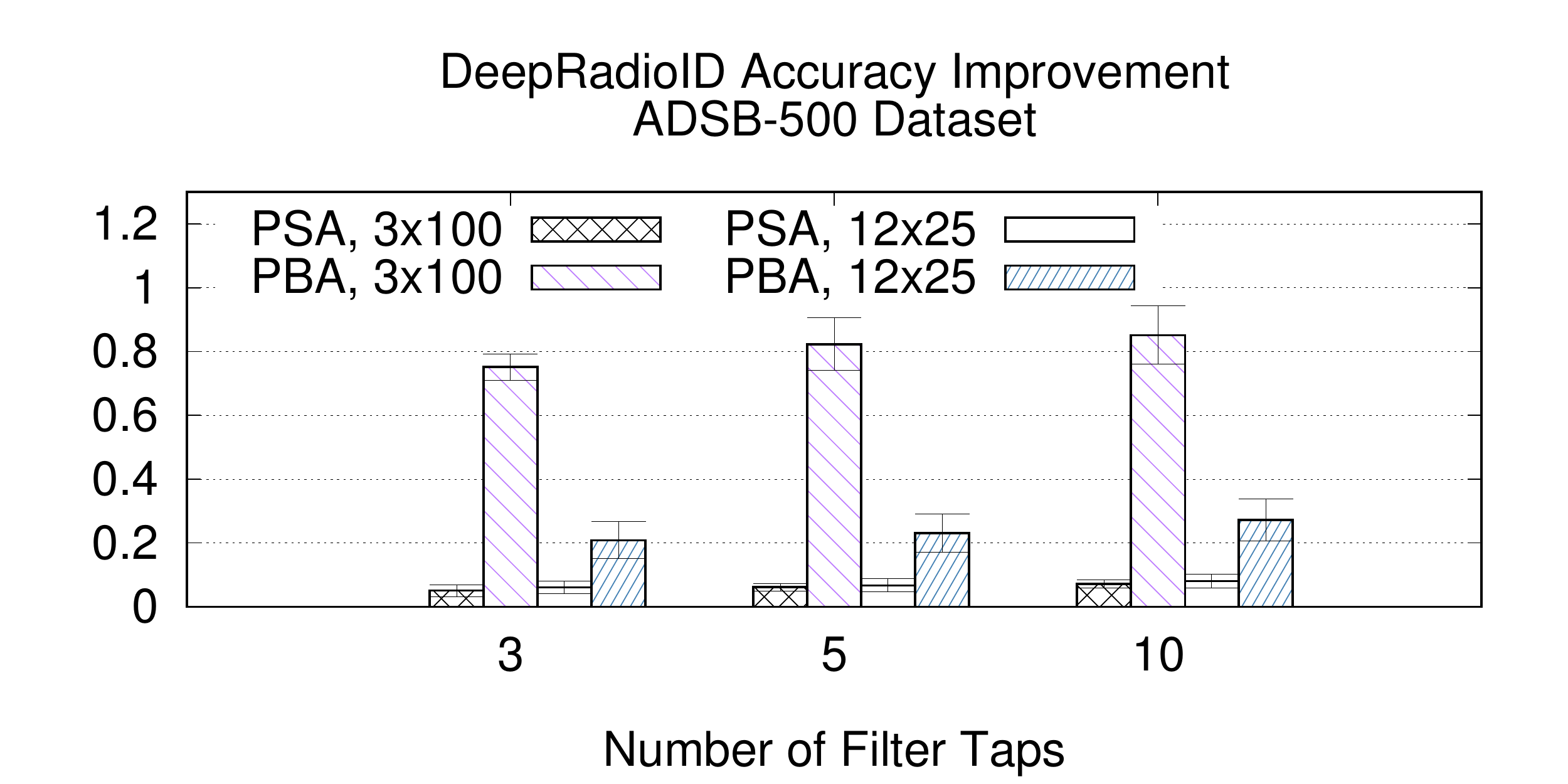}
    \caption{Per-Slice and Per-Batch Accuracy Improvement, ADSB-500, as a function of the number of inputs per slice (100 and 25) and the number of batches (3 and 12).\vspace{-0.3cm}}
    \label{fig:adsb500improvement}
\end{figure}

Figure \ref{fig:adsb500improvement} reports the PSA and PBA improvement on the ADS-B-500 dataset (baseline PSA and PBA of .5028 and .6193), which reach respectively about 10\%  and 50\% in the case of 10 filter taps. Figure \ref{fig:adsb500improvement} also shows that,  although the ADSB-500 PSA improvement is about the same experienced in the WiFi-100 model, the PBA improvement is significantly higher in ADSB-500 (30\% vs 50\%). This is thanks to the fact that a very small increase in PSA usually corresponds to a significant increase in PBA when the number of devices is higher. Also, ADS-B transmits in a less-crowded channel than WiFi, thus we expect our model and \textit{DeepRadioID}
to perform better on ADSB-500 when the number of inputs per slice increases.

\begin{figure}[!h]
    \centering
    \includegraphics[width=0.5\columnwidth,angle=-90]{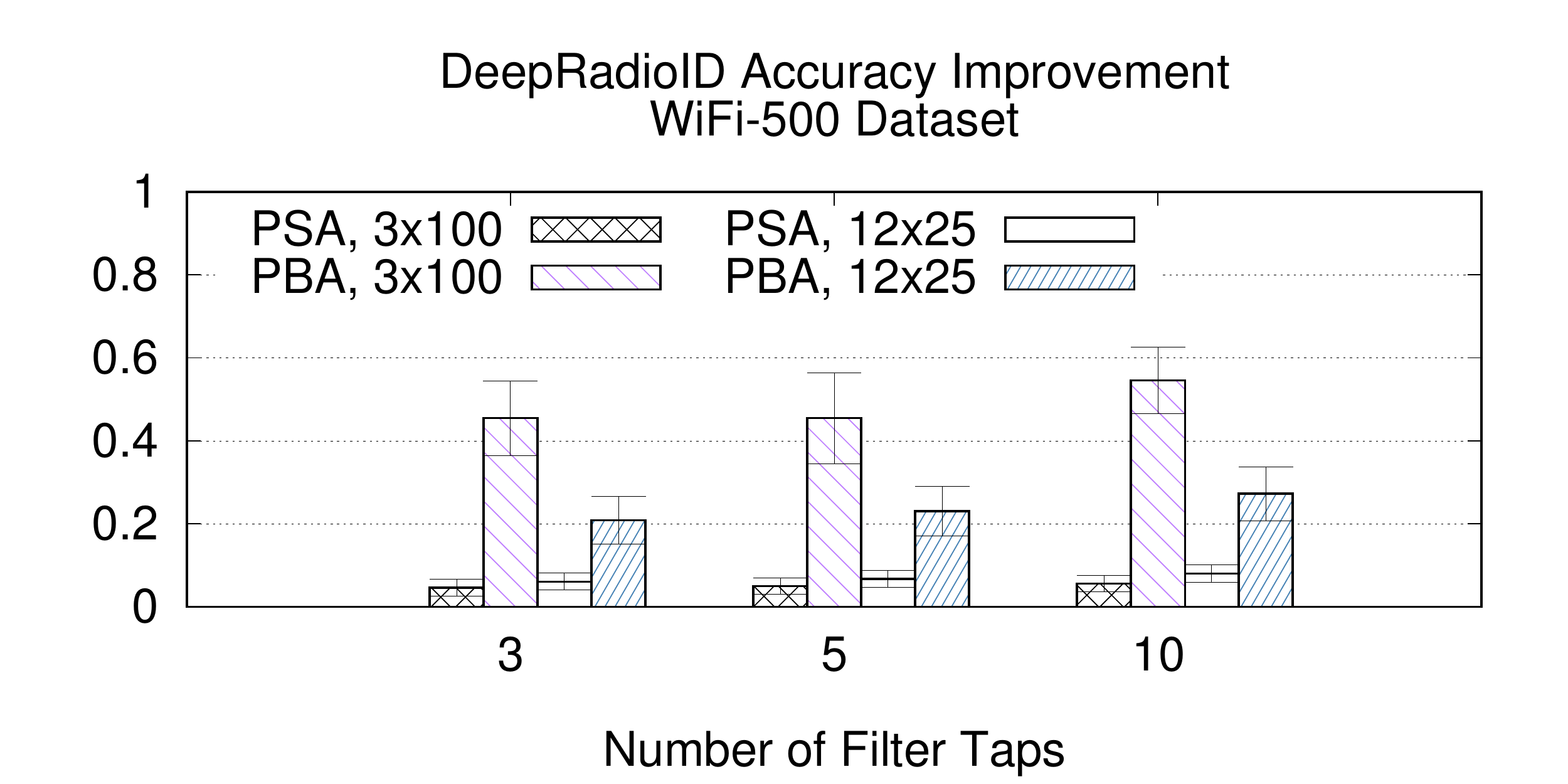}
    \caption{Per-Slice and Per-Batch Accuracy Improvement, WiFi-500, as a function of the number of inputs per slice (100 and 25) and the number of batches (3 and 12).\vspace{-0.3cm}}
    \label{fig:wifi500improvement}
\end{figure}

This is also confirmed by Figure \ref{fig:wifi500improvement}, where we show the PSA and PBA improvement in the WiFi-500 dataset as a function of the number of FIR taps and the number of inputs in each slice (25 and 100). Indeed, Figure \ref{fig:wifi500improvement} shows a similar PBA improvement when 10 FIR taps are considered. Furthermore, Figure \ref{fig:wifi500improvement} shows that the number of inputs per slice significantly impacts on the fingerprinting accuracy, especially on the PBA, in both ADSB-500 and WiFi-500. This is because (i) the FIR optimization increases in effectiveness as the number of inputs per slice increases, because the FIR is averaged over more channel realizations; and (ii) the boosting effect given by the PBA increases as the number of slices increases.

\begin{figure}[!h]
    \centering
    \includegraphics[width=0.5\columnwidth,angle=-90]{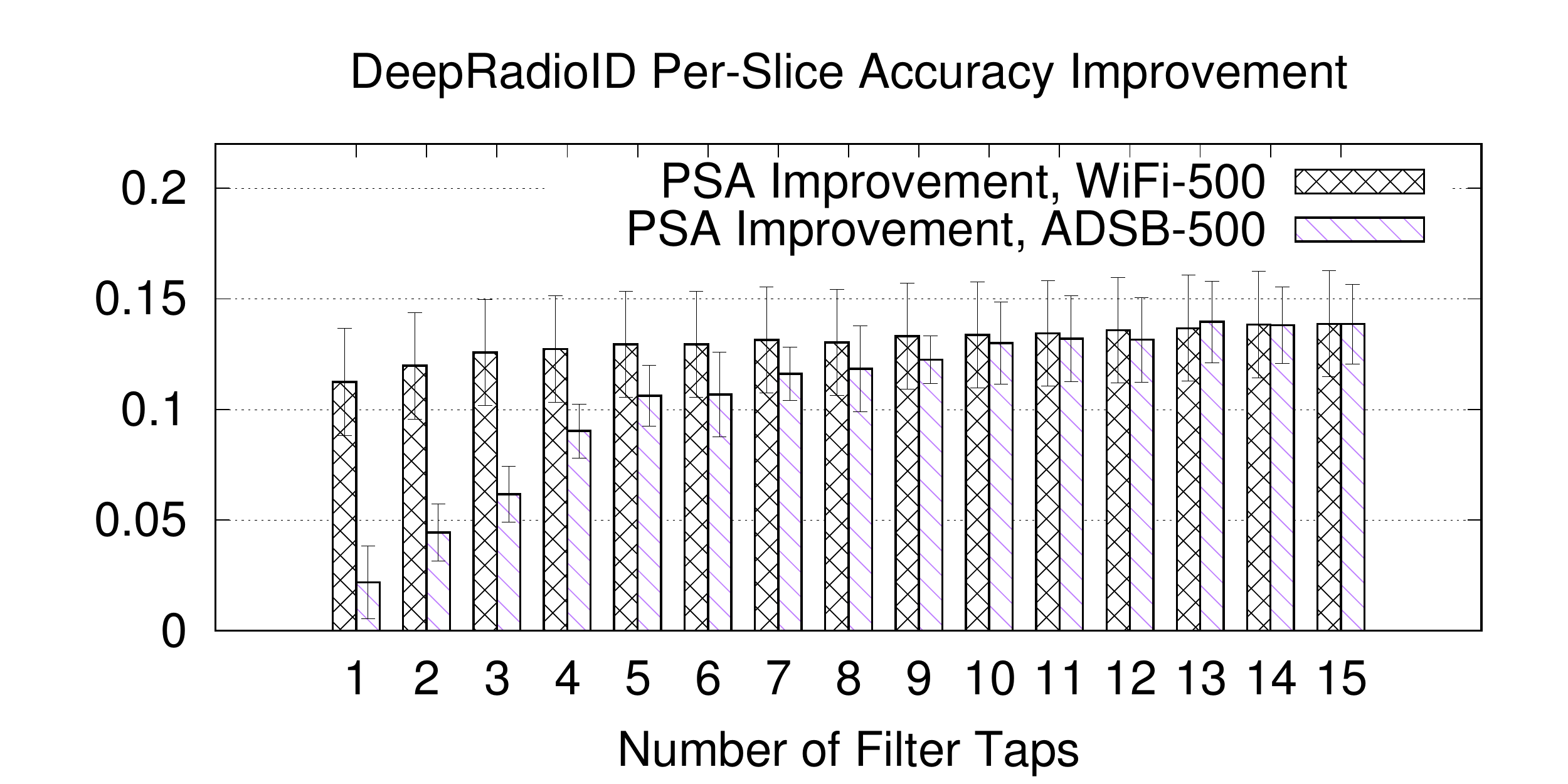}
    \caption{PSA Improvement as a function of the number of FIR taps. ADSB-500 converges slower than WiFi-500 given the FIR has to operate on unprocessed I/Q samples. \vspace{-0.3cm}}
    \label{fig:filtertapsadsb500}
\end{figure}

Figure \ref{fig:filtertapsadsb500} shows the PSA improvement for ADSB-500 and WiFi-500 as a function of the number of FIR taps. As we can see, the PSA improvement converges to approximately .15 as more FIR taps are used, in both cases. Very interestingly,  Figure \ref{fig:filtertapsadsb500} also shows that WiFi-500 converges more rapidly than ADSB-500. This is due to the fact that our ADSB-500 model was trained on unprocessed I/Q samples (\textit{i.e.}, without any channel equalization). Therefore, the number of FIR taps than \emph{DeepRadioID} needs to obtain the same performance as in WiFi-500 increases, as the FIR has to compensate for a significant channel action already mitigated in WiFi-500.

\begin{figure}[!h]
    \centering
    \includegraphics[width=0.5\columnwidth,angle=-90]{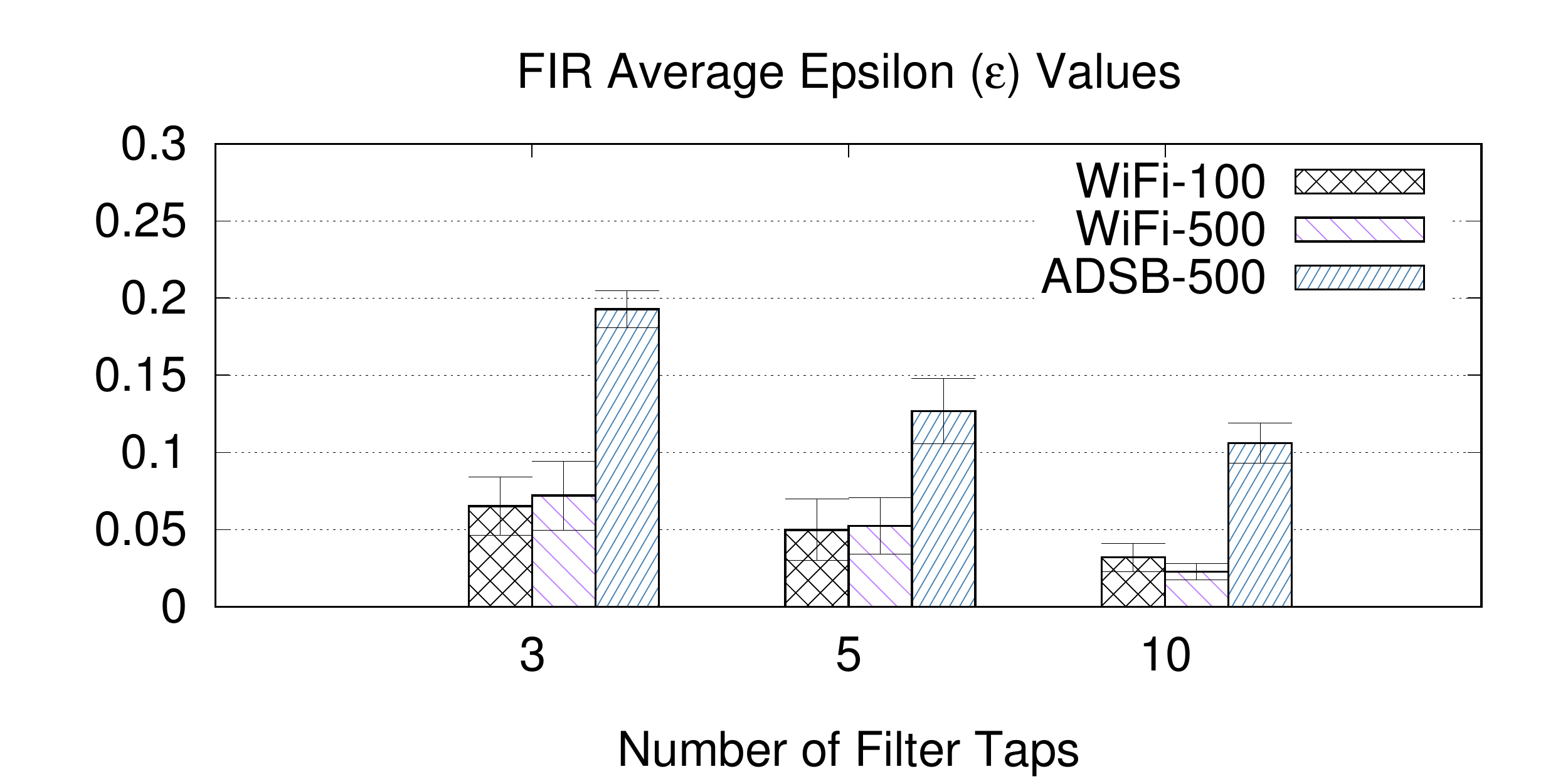}
    \caption{Average FIR Epsilon ($\epsilon$) values as a function of the number of FIR taps.\vspace{-0.3cm}}
    \label{fig:epsilon_taps}
\end{figure}

Finally, Figure \ref{fig:epsilon_taps} depicts the average FIR $\epsilon$-value (defined in Section \ref{sec:ber}) as a function of the number of taps and the model considered. As anticipated in Section \ref{sec:ber}, Figure \ref{fig:epsilon_taps} concludes that the maximum $\epsilon$-value is below .2, which allows us to achieve low BER even when the FIR is applied. Interestingly enough, we notice that ADSB-500 requires taps that are on average higher than WiFi-100 and WiFi-500. This is due to the fact that ADSB-500 operates on unprocessed I/Q samples in the time domain, therefore the taps required to modify the signal need to be greater.

\section{Conclusions}

In this paper we have proposed \emph{DeepRadioID}, a system to optimize the accuracy of deep-learning-based radio fingerprinting algorithms. We have extensively evaluated \emph{DeepRadioID} on a experimental testbed of 20 nominally-identical software-defined radios, as well as on datasets made up by WiFi and ADS-B transmissions. Experimental results have shown that \emph{DeepRadioID} (i) increases fingerprinting accuracy by about 35\%, 50\% and 58\% on the three scenarios considered; (ii) decreases an adversary's accuracy by about 54\% when trying to imitate other device's fingerprints by using their filters; (iii) achieves 27\% improvement over the state of the art on a 100-device dataset.

\section*{Acknowledgment}

This work is supported by the Defense Advanced Research Projects Agengy (DARPA) under RFMLS program contract N00164-18-R-WQ80. We are sincerely grateful to Paul Tilghman, program manager at DARPA, Esko Jaska, our shepherd Srinivas Shakkottai and the anonymous reviewers for their insightful comments and suggestions. The views and conclusions contained herein are those of the authors and should not be interpreted as necessarily representing the official policies or endorsements, either expressed or implied, of DARPA or the U.S. Government.




\end{document}